\documentclass[twocolumn,amsmath,amssymb,nofootinbib]{revtex4}
\usepackage{amsmath}
\usepackage{natbib}
\def\be{\begin{equation}}
\def\ee{\end{equation}}
\def\ba{\begin{eqnarray}}
\def\ea{\end{eqnarray}}
\def\bdm{\begin{displaymath}}
\def\edm{\end{displaymath}}
\def\bwt{\begin{widetext}}
\def\ewt{\end{widetext}}
\def\be{\begin{equation}}
\def\ee{\end{equation}}
\def\ba{\begin{eqnarray}}
\def\ea{\end{eqnarray}}

\def\bdm{\begin{displaymath}}
\def\edm{\end{displaymath}}

\def\bq{\begin{quote}}
\def\eq{\end{quote}}

 at 10truept

\newcommand{\R}{{\cal R}}
\newcommand{\beq}{\begin{equation}}
\newcommand{\eeq}{\end{equation}}
\newcommand{\beqa}{\begin{eqnarray}}
\newcommand{\eeqa}{\end{eqnarray}}

\begin{document}

\title{Rotating Black Holes on Codimension-2 Branes}

\author{Derrick Kiley}
\email{dtkiley@physics.ucdavis.edu}
\affiliation{Department of Physics, University of California,
Davis,
CA 95616}

\date{\today}

\begin{abstract}
It has recently been demonstrated that certain types of non-tensional
stress-energy can live on tensional codimension-2 branes, including
gravitational shockwaves and small Schwarzschild black holes.  In this note we generalize the earlier Schwarzschild results,
and construct the exact gravitational fields of small rotating black
holes on a codimension-2 brane.  We focus on the phenomenologically interesting
case of a three-brane embedded in a spacetime with two compactified
extra dimensions.  For a nonzero tension on the brane, we verify that
these solutions also show the ``lightning rod'' effect found in the
Schwarzschild solutions, the net effect of which is to rescale the fundamental Planck mass. This allows for larger black hole parameters, such as the event horizon, angular momentum, and
lifetime than would be naively expected for a tensionless brane.   It is also found that a black hole with  angular momentum pointing purely along the brane directions has a smaller horizon angular velocity than the corresponding tensionless case, while a hole with bulk components of angular momentum has a larger angular velocity.

\end{abstract}

\maketitle

\section{Schwarzschild Black Holes on Codimension-2 Branes}
It has long been known that the fundamental scale of gravity could
be much lower than the four-dimensional Planck scale, $M_P \sim
10^{19}$ GeV in theories with extra dimensions
\cite{Arkani-Hamed:1998rs}-\cite{Dvali:2000hr}.  This realization
has generated considerable interest in the possibility that
microscopic black holes can form at energies that will be achievable
at the forthcoming Large Hadron Collider (LHC). If the fundamental
gravitational scale, $M_*$, is of the order of a TeV, then the LHC
may be able to create these black holes through a gravitational
lensing process \cite{Kaloper:2007pb} at the rate of approximately
one per second \cite{Giddings:2001bu}, which then quickly decay
through the emission of Hawking radiation.  This would provide a
very interesting arena in which one can study quantum gravity
effects and has been investigated extensively in the literature, for
example in \cite{Giddings:2001bu}-\cite{Chen:2007jz}.  One
particularly interesting avenue along which one can study small
black holes is to put them on codimension-2 branes in which a
3-brane floats in a $6D$ bulk, for example.

Codimension-2 branes have been the topic of much interest recently
\cite{Sundrum:1998ns}-\cite{Kobayashi:2007hf} because of their
remarkable properties.  In general, putting a brane with tension in a
bulk spacetime curves the brane as well as the bulk, making the system typically
very difficult to analyze. However, on
codimension-2 branes the vacuum energy on the brane can be
``offloaded'' into the bulk cusping it into a cone centered on the
brane, leaving the brane flat.  This considerably simplifies the task of
finding solutions. For example, exact solutions have been found in
\cite{Kaloper:2006ek} for a relativistic particle. In the
relativistic limit the equations of motion of the particle are exactly described by
linear equations and yield a gravitational shockwave, generalizing the $4D$ Aichelburg-Sexl solutions \cite{Aichelburg:1970dh}. These
shockwave solutions can be viewed as a six-dimensional black hole on
the brane boosted to a relativistic speed, since the gravitational
field lines are flattened completely transverse to the direction of
motion due to Lorentz contraction.  The exact Schwarzschild black
hole solution on a tensional codimension-2 brane is, to our
knowledge, the \emph{first} example of an exact black hole on a
3-brane, which we now review.

The black hole is constructed in analogy with the method of Aryal,
Ford, and Vilenkin (AFV) \cite{Aryal:1986sz}, which describes a
black hole threaded by a straight cosmic string of mass per unit
length $\mu$. The net effect of the cosmic string is to induce a
conical deficit in the surrounding spacetime, which remains locally
flat, but has a conical singularity along the string.  The AFV
solution is constructed by starting with the four-dimensional
Schwarzschild solution.  The cosmic string is then threaded along an
axis of symmetry of the black hole (which is trivial in the
Schwarzschild case).  One accounts for its presence by cutting
out a wedge from the polar angle around the symmetry axis, the size
of which is proportional to $\mu$, and then identifying the edges of
the cut, rescaling the polar angle, $\phi \rightarrow b\phi$, where $b
= 1-4\mu$ \cite{Aryal:1986sz}. This gives the conical topology since the polar angle
doesn't run around a full $2\pi$ radians.

A small black hole on a tensional 3-brane is constructed along the
same lines, where by ``small'' we mean that the hole has a horizon
size smaller than the compactification scale of the extra dimensions, making it a true
six-dimensional object.  Here we follow the construction given in
\cite{Kaloper:2006ek}.  We start with the six-dimensional Schwarzschild solution
given by
\be ds_6^2 =
-\left(1-\left(\frac{r_0}{r}\right)^3\right)dt^2+\frac{dr^2}{1-\left(\frac{r_0}{r}\right)^3}+r^2
d\Omega_4,\\
\label{eqn:6Dsch} \ee
where $r_0$ is the event horizon which depends on the mass of the
black hole, and $d\Omega_4$ is the metric of a $4D$ unit sphere.  To
include the brane, we choose a three-dimensional symmetry hypersurface along which we
thread the brane (generalizing the axis of symmetry in the AFV solution).  Then, including the tension on the brane tells us
that we should cut out a wedge from the polar angle that runs around
that hypersurface, the size of the wedge depending on the tension, $\lambda$.

The method discussed above is simple in the so-called uniform
coordinates, obtained from Eq. (\ref{eqn:6Dsch}) by the coordinate
transformation
\be r = \R\left(1+\frac{1}{4}\left(\frac{r_0}{\R}\right)^3\right)^{\frac{2}{3}},\\
\label{eqn:6Dschdiff} \ee
in terms of which Eq. (\ref{eqn:6Dsch}) becomes
\be
\begin{array}{ll}
ds_6^2 = -\left(\frac{4
\R^3-r_0^3}{4\R^3+r_0^3}\right)dt^2\\
\quad \quad \quad \quad +\left(1+\frac{1}{4}\left(\frac{r_0}{\R}\right)^3\right)^{\frac{4}{3}}\left(d\R^2+\R^2
d\Omega_4\right).\\
\end{array}
\label{eqn:6Dschiso} \ee
We can write the metric on the unit sphere in terms of polar
coordinates, $\R^2 = \vec{x}^2+\rho^2$, where $\vec{x}$ denotes the
coordinates along the brane, and $\rho$ measures distance in the
bulk.  Then $d\R^2 +
\R^2 d\Omega_4 = d\vec{x}^2+d\rho^2 + \rho^2 d\phi^2$. Finally, to include the brane we cut out a wedge in the $\rho,\phi$
plane of angular opening $\lambda/M_6^4$, then rescale $\phi \rightarrow
B\phi$, where \cite{Kaloper:2006ek}
\be
B = 1-\frac{\lambda}{2\pi M_6^4}.\\
\label{eqn:B}
\ee
This gives the exact solution for a Schwarzschild black hole on a
tensional 3-brane
%
\be
\begin{array}{ll}
ds_6^2 = -\left(\frac{4
\R^3-r_0^3}{4\R^3+r_0^3}\right)dt^2\\
\quad \quad \quad \quad +\left(1+\frac{1}{4}\left(\frac{r_0}{\R}\right)^3\right)^{\frac{4}{3}}\left(d\vec{x}^2+d\rho^2
+ B^2\rho^2 d\phi^2\right).\\
\end{array}
\label{eqn:6DschisoBpolar} \ee
%

While the above construction was performed in uniform coordinates to
facilitate easy comparison with the AFV solution, one can easily
express the metric in a Schwarzschild form.  In this case one starts
with Eq. (\ref{eqn:6Dsch}) and writes $d\Omega_4 = d\Omega_3 + B^2
\prod_{k = 1}^3 \sin^2(\theta_k)d\psi^2$, which gives \cite{Dai:2006hf}
%
\be
\begin{array}{ll}
ds_6^2 = -\left(1-\left(\frac{r_0}{r}\right)^3\right)dt^2 +
\frac{dr^2}{1-\left(\frac{r_0}{r}\right)^3} + r^2\big\{d\theta^2\\
\quad \quad \quad \quad +\sin^2\theta\left[d\phi^2 + \sin^2\phi\left(d\chi^2 + B^2 \sin^2 \chi\,
  d\psi^2\right)\right]\big\},\\
\end{array}
\label{eqn:6DBsch}
\ee
%
which is particularly simple, and one can check that it still
satisfies Einstein's equations in the bulk.  It's easy to see that
Eq. (\ref{eqn:6DBsch}) can be obtained from Eq. (\ref{eqn:6Dsch})
simply by rescaling the angle $\psi \rightarrow B\psi$.  It is further clear that $\psi$ has associated with it a Killing
vector, $\partial_\psi$, which immediately suggests that $\psi$ is the angle to
rescale\footnote{Notice that the same argument holds for
  Eq. (\ref{eqn:6DschisoBpolar}), where $\partial_\phi$ is the Killing
  vector, suggesting that we rescale $\phi$ in that form of the metric.}.  This will be useful when we extend this construction
to the rotating black hole solutions.

The ADM mass of the black hole is \cite{Kaloper:2006ek}
\be
m = 2 M_6^4 r_0^3\int\,d\Omega_4.\\
\label{eqn:ADMmass}
\ee
Performing the integral in the tensionless case yields the area of the
4-sphere.  However, when we include a nonzero $\lambda$, the
full integration is over a sphere with a deficit angle, and so the
area element depends on the tension, $d\Omega_4 \rightarrow B d\Omega_4$.  The integral in
Eq. (\ref{eqn:ADMmass}) gives a factor of
$2\pi^2 B$, and so
\bdm
m = 4\pi^2 B M_6^4 r_0^3,\\
\edm
from which we can read off the horizon size
\be
r_0 = \left(\frac{m}{4\pi^2 \,BM_6^4}\right)^{\frac{1}{3}} \equiv \frac{1}{B^{1/3}}r_S,\\
\label{eqn:6Dschhorizon}
\ee
where $r_S$ is the usual $6D$ Schwarzschild radius in the zero tension
limit.

Eq. (\ref{eqn:6Dschhorizon}) shows a very interesting feature.  Since
$B = 1-\frac{\lambda}{2\pi M_6^4}$ we see that the horizon size can be
much larger than naively expected on the basis of the black hole's
mass alone.  In particular, the horizon size can become very large
for near-critical branes where $\lambda \rightarrow 2\pi M_6^4$. Of
  course, the horizon size cannot grow too large, as the fundamental
  description of the hole as a $6D$ object begins to fail as the horizon size
  approaches scales comparable to the compactification radius of the
  extra dimensions.  After
  that, the hole will look approximately four-dimensional, where the
  $4D$ Planck mass is determined in the usual way using Gauss's law, $M_4^2 = M_6^4
  \times \textrm{Vol}(y^a)
  \sim (B L^2)M_6^4$, where $L$ is the compactification radius.  Thus,
  the $4D$ Planck scale already includes the effects of a non-zero
  tension.

Qualitatively, the reason for this enhancement of the horizon is clear: the extra
dimensions are bent into a cone, and so the gravitational field
lines cannot spread out as quickly as would be the case in a full
spherical geometry.  The presence of the tension on the brane has
caused an amplification of the gravitational field, behaving in effect
like a ``lightning rod'' for gravity, in much the same way as the
electric field can become very large around a charged needle.  In
turn, supercritical tensions compactify the bulk on a $2D$ teardrop,
changing gravity to a lower-dimensional one
\cite{Kaloper:2007my,Kaloper:2007ap}.  In what follows we will assume that the tension is safely
\emph{non-critical}, $\lambda <2\pi M_6^4$, although we allow the
tension to still be large
enough to cause significant enhancement.

If we rewrite $B M_6^4 \equiv M_6^4{}_\textrm{eff}$ in
Eq. (\ref{eqn:6Dschhorizon}) then it is obvious that the net effect of
the tension on the brane is to rescale the six-dimensional Planck
mass.  Since the gravitational coupling goes $\sim
1/M_6^4{}_\textrm{eff}$, we see that the effective coupling is larger
for a nonzero tension than that set by simply the fundamental
Planck scale, $M_6$.  This explains why the horizon size can be larger
than expected -- gravity appears stronger.  We will also verify this lightning rod effect when we
extend the black hole solutions to include spin, to which we turn now.

\section{Rotating Black Holes on Codimension-2 Branes}

The metric for a spinning black hole in six dimensions is considerably
more complicated than the simple Schwarzschild solution given in
Eq. (\ref{eqn:6Dsch}), and can be expressed in Boyer-Lindquist coordinates as
in \cite{Myers:1986un}. Here we differ slightly from
  \cite{Myers:1986un} and take the opposite sign for the
  two angular momentum parameters, $a_i \rightarrow -a_i$ to be more
  consistent with other literature.  Note, also, that $r$ is
  \emph{not} the usual flat-space radial coordinate, but is defined
  instead by the elliptical coordinate relation $\frac{(x^1)^2+(x^2)^2}{r^2 +
    a_1^2} + \frac{(y^1)^2 + (y^2)^2}{r^2 + a_2^2} + \frac{(x^3)^2}{r^2} =1$,
  where $x^i$ are the brane coordinates, and $y^a$ are the bulk coordinates.
 With this the metric is
\be
\begin{array}{ll}
ds_6^2 = -dt^2 + r^2 d\alpha^2 + \frac{\Pi F}{\Pi - \mu r}dr^2 +\\
\quad \quad \sum_{i=1}^2\left[ (r^2 + a_i^2)\big(d\mu_i^2 + \mu_i^2
d\phi_i^2\big) + \frac{\mu r}{\Pi F}\big(dt - a_i \mu_i^2 d\phi_i\big)^2 \right].\\
\end{array}
\label{eqn:muimetric}
\ee
%
Here $\mu$ is proportional to the mass of the black hole, and
\be
\Pi \equiv \prod_{i = 1}^2 (r^2 + a_i^2),\\
\label{eqn:muiPi}
\ee
\be
F \equiv \sum_{i=1}^2 \frac{r^2 \mu_i^2}{r^2 + a_i^2} + \alpha^2.\\
\label{eqn:muiF}
\ee
The two direction
cosines, $\mu_i$, and the $\alpha$ coordinates are subject to the
constraint
\be
\mu_1^2+\mu_2^2 + \alpha^2 = 1,\\
\label{eqn:mualphaconstraint}
\ee
which could be satisfied by expressing $\mu_i$ and $\alpha$ in terms of
angles on a unit $2-$sphere, for example, but we'll leave the expressions general for now.  In general a $D-$dimensional Kerr solution has $\left \lfloor
\frac{D-1}{2}\right \rfloor$ angular momentum parameters, where $\left
\lfloor x\right \rfloor$ denotes the integer part of $x$.  Hence,
Eq. (\ref{eqn:muimetric}) has two parameters, $a_i$.  Unlike in four
dimensions, for six or more dimensions black holes with a fixed mass
can have an arbitrarily large angular momentum \cite{Myers:1986un}.

Eq. (\ref{eqn:muimetric}) is written in a compact form, but in order to extend
the spinning black hole solutions to include a nonzero tension, it
will be more convenient to expand the sum in
Eq. (\ref{eqn:muimetric}), rewriting it in the following way
\bwt
\be
\begin{array}{l}
ds_6^2 = -\left(1-\frac{\mu r}{\Pi F}\right) dt^2 - \frac{2\mu r}{\Pi
  F}\left(a_1\mu_1^2 d\phi_1 + a_2\mu_2^2 d\phi_2\right)dt +
   \frac{F}{1-\frac{\mu \,r}{\Pi}}dr^2+r^2\left(d\alpha^2 + d\mu_1^2 + d\mu_2^2 + \mu_1^2 d\phi_1^2 +
  \mu_2^2 d\phi_2^2\right)\\
\quad \quad \quad \quad + a_1^2\left[d\mu_1^2 + \mu_1^2\left(1+\frac{\mu r}{\Pi
    F}\mu_1^2\right)d\phi_1^2\right] + a_2^2 \left[d\mu_2^2 + \mu_2^2\left(1+\frac{\mu r}{\Pi F}\mu_2^2\right)d\phi_2^2\right]
  + 2a_1 a_2\frac{\mu r}{\Pi F}\mu_1^2 \mu_2^2 d\phi_1 d\phi_2.\\
\end{array}
\label{eqn:abmetric}
\ee
\ewt

To include the brane we proceed as before and choose a symmetry
hypersurface and thread the brane along it.  Because we have two
angular momentum parameters we now have a choice of two hypersurfaces
along which we can thread the brane.  We can choose the normal to the
plane swept out by either $\phi_1$ or $\phi_2$.  Once again, it's
  clear that the Killing vectors, $\partial_{\phi_1}$,
  $\partial_{\phi_2}$ suggest the angles that we should rescale.  The general metric,
Eq. (\ref{eqn:abmetric}), is
symmetric under the simultaneous interchange $a_1 \leftrightarrow a_2$ and $\phi_1
\leftrightarrow \phi_2$.  So without loss of generality, let us choose
$\phi_2$ as defining the angle about the symmetry hypersurface.  Then, rescaling
$\phi_2 \rightarrow B \phi_2$ as in Eq. (\ref{eqn:6DBsch}) yields our metric
\bwt
\be
\begin{array}{l}
ds_6^2 = -\left(1-\frac{\mu r}{\Pi F}\right) dt^2 - \frac{2\mu r}{\Pi
  F}\left(a_1\mu_1^2 d\phi_1 + a_2B\mu_2^2 d\phi_2\right)dt +
  \frac{F}{1-\frac{\mu \,r}{\Pi}}dr^2 + r^2\left(d\alpha^2 + d\mu_1^2 + d\mu_2^2 + \mu_1^2 d\phi_1^2 +
  B^2\mu_2^2 d\phi_2^2\right)\\
\quad \quad \quad \quad + a_1^2\left[d\mu_1^2 + \mu_1^2\left(1+\frac{\mu r}{\Pi
    F}\mu_1^2\right)d\phi_1^2\right] + a_2^2 \left[d\mu_2^2 + B^2 \mu_2^2\left(1+\frac{\mu r}{\Pi F}\mu_2^2\right)d\phi_2^2\right]
+ 2a_1B a_2\frac{\mu r}{\Pi F}\mu_1^2 \mu_2^2 d\phi_1 d\phi_2.\\
\end{array}
\label{eqn:abBmetric}
\ee
\ewt

Eq. (\ref{eqn:abBmetric}) has an interesting feature.  The obvious
choice for orienting the symmetry hypersurface would be the plane defined by the
axis of rotation of the black hole.  This would be the only choice in
the analogous construction of a spinning black hole threaded by a
cosmic string.  However, in Eq. (\ref{eqn:abBmetric}) we have two
choices of angle through which the hole can rotate. We have chosen to rescale $\phi_2 \rightarrow
B\phi_2$, orienting the brane along the normal to $\phi_2$, but we have not specified the rotation axis.  Choosing the
rotation axis to be $\phi_2$, by setting $a_1 \equiv 0$, say,
corresponds to cutting the wedge out of the rotation axis such that
the hole would not complete a full revolution through $2\pi$ radians,
but only through $2\pi B$ radians.  The hole would be spinning orthogonal to the brane with its angular momentum pointing along the brane. Choosing
$\phi_1$ instead cuts the wedge from an axis orthogonal to the
direction of spin, such that the hole still completes a full $2\pi$
radian revolution.  The hole would be spinning on the brane with its angular momentum orthogonal to the brane.

Eq. (\ref{eqn:abBmetric}) again demonstrates the lightning rod
effect, as we can see by determining $\mu$.  To leading order, the
mass of the black hole, $M$, is related to the metric perturbation in six
dimensions by \cite{Myers:1986un}
\be
h_{00}\approx \frac{M}{2 A_4(B) M_6^4}\frac{1}{r^3},\\
\label{eqn:h00mass}
\ee
where $A_4(B)$ is the area of the unit $4-$sphere, rescaled to include
the conical singularity.  At asymptotically large distances from the black hole the metric
in Eq. (\ref{eqn:abBmetric}) contains $g_{00} \approx -1 +
\frac{\mu}{r^3}$, and so comparison with Eq. (\ref{eqn:h00mass})
gives
\be
\mu = \frac{M}{4\pi^2 M_6^4{}_\textrm{eff}},\\
\label{eqn:mu}
\ee
after plugging in $A_4(B) = 2\pi^2 B$, and redefining $BM_6^4 \equiv M_6^4{}_\textrm{eff}$.  It
is unsurprising that the lightning rod effect appears here, as this
aspect of the rotating solution exactly follows the Schwarzschild
case.

While the Schwarzschild black hole is completely characterized
by its mass, the rotating black hole solution is also characterized
by its spin, and we might expect that the lightning rod effect may
play a role there, as well.  The angular momentum, given in terms of
the parameters $a_i$ \cite{Myers:1986un} 
\be
J_i = A_4(B) M_6^4 \mu a_i = \frac{1}{2}M a_i,\\
\label{eqn:angmomai}
\ee
where $M$ is the black hole mass, does not appear to change since the
tension cancels out after plugging in Eq. (\ref{eqn:mu}) for $\mu$.
However, closer inspection reveals that the angular momentum for a fixed mass can also be larger than expected, as we will see
below.

\section{A Special Case -- A Single Rotation Parameter}

It is already clear that the tension on the brane can have an
interesting effect on the black hole properties through the lightning
rod effect.  We will see that this effect will persist to other aspects of
the black hole such as the temperature and the lifetime.  The metric in Eq. (\ref{eqn:abBmetric}) is general, as are all the
expressions derived from it so far and we could continue with the most
general case.  In this section, however, let us focus on a
particular case of phenomenological interest that will elucidate the
effects of the tension in as simple and direct a way as possible.

We suppose that the
black hole is formed through the collision of two particles on a delta
function brane.  If the collision occurs with a nonzero impact
parameter, $b$, then the hole will have angular momentum.  Since the
particles are constrained to live on the brane, the hole will spin on the brane and the initial
angular momentum will have a single value, directed along the brane.
By conservation of angular momentum, the black hole will subsequently
have a single nonzero rotation parameter, $a$.  We can determine a
simpler form for the metric in this limit starting with
Eq. (\ref{eqn:abBmetric}) and setting $a_2 \equiv 0$ and letting $a_1
= a$, which is required by the combination of our choice of rescaling $\phi_2 \rightarrow
  B\phi_2$, and also having the angular momentum point along the brane.  Then making the substitutions $\phi_1 = \phi$, $\phi_2 = \psi$, and
\bdm
\mu_1 = \sin \theta, \quad \mu_2 = \cos \theta \, \sin \chi, \quad
\alpha = \cos \theta\, \cos \chi,\\
\edm
we find
\bwt
\be
\begin{array}{ll}
ds_6^2  = & -\left(1-\frac{\mu}{r\rho^2}\right)dt^2 - \frac{2\mu
  a}{r\rho^2}\sin^2\theta\,dt d\phi + \frac{\rho^2}{r^2 + a^2 -
  \mu/r}dr^2\\
& \quad \quad + \rho^2 \,d\theta^2
 +\sin^2\theta\left(r^2 + a^2 + \frac{\mu
  a^2}{r\rho^2}\sin^2\theta\right)d\phi^2 + r^2 \cos^2\theta
  \left(d\chi^2 + B^2\sin^2\chi \,d\psi^2\right).\\
\end{array}
\label{eqn:aBmetric}
\ee
\ewt
where $\rho^2 = r^2 + a^2 \cos^2\theta$.  Let us now analyze the
parameters associated with Eq. (\ref{eqn:aBmetric}).  Eq. (\ref{eqn:mu}) still
gives $\mu$, while Eq. (\ref{eqn:angmomai}) reduces to
\be
J = \frac{1}{2}Ma.\\
\label{eqn:angmoma}
\ee

The particles have a center of mass energy $\sqrt{s} = M_i$, which is
fixed by the accelerator, with an impact parameter $b$.
Then, the initial angular momentum will be $J_i = b M_i/2$ in the
center of mass frame.  The black hole is formed when the particles
pass within a distance of each other smaller than the event horizon
of the black hole which depends
on its mass and also its angular momentum.  This sets
an upper limit on the impact parameter of $b \le 2r_H(M,J)$ \cite{Ida:2002ez}, and so
therefore an upper limit on the angular momentum (for a fixed mass).
After the collision, a black hole of mass $M$ and
angular momentum $J = b M/2$ is formed.  Comparison with
Eq. (\ref{eqn:angmoma}) shows that $a = b$, and so the angular
momentum parameter is really just a measure of the impact parameter
(for a black hole formed in this way).  Plugging in for $a \rightarrow b_\textrm{max}$ in
Eq. (\ref{eqn:angmoma}) we find
\be
J_\textrm{max} = \frac{M b_\textrm{max}}{2} = M r_H.\\
\label{eqn:Jmax}
\ee

The horizon occurs where $r_H^3 + r_H a^2 =
\mu$, which we can write as
\bdm
r_H^3\left(1+\frac{a^2}{r_H^2}\right) \equiv r_H^3\left(1+a_*^2\right)
= \mu,\\
\edm
where $a_* \equiv a/r_H = 2J/J_\textrm{max}$.  Because $0 \le a_* \le 2$, we have $\frac{1}{5}\mu \le r_H^3 \le \mu$, so
$a_*$ makes very little difference and we can take the above
expression as defining the event horizon of the spinning black
hole rather than solving the cubic equation.  Then, using Eq. (\ref{eqn:mu}),
\be
r_H = \left(\frac{M}{4\pi^2(1+a_*^2)M_6^4{}_\textrm{eff}}\right)^{\frac{1}{3}}.\\
\label{eqn:horizon}
\ee
Note the presence of the
effective Planck scale in Eq. (\ref{eqn:horizon}), showing the lightning
rod effect.  It's immediately clear that the angular momentum, Eq. (\ref{eqn:Jmax}), can be larger than would be
naively expected ($\sim B^{-1/3}J_{0\textrm{max}}$, where $J_{0\textrm{max}}$ is the tensionless
case).  It is also clear that the ergosphere will be similarly enhanced.

Given that for a black hole of a fixed mass the horizon can be larger for a tensional brane 
than for one without tension, one would also expect the
temperature of the black hole to decrease as a function of mass since large black holes have
a lower temperature than small ones.  
The temperature can be
determined by Wick rotating the time coordinate in the usual way
and is given
by
\be
T_H =\frac{3+a_*^2}{4\pi r_H(1+a_*^2)},\\
\label{eqn:temp}
\ee
which is, indeed, smaller than the zero tension limit.
Eq. (\ref{eqn:temp}) has an associated entropy
\be
S_{BH} = \frac{M}{4T_H}\left(\frac{3+a_*^2}{1+a_*^2}\right),\\
\label{eqn:entropy}
\ee
and is also
larger than one would naively expect.

Following their production, the black holes evaporate first shedding
their hair during a ``balding'' phase.  The holes then enter a
spin-down phase during which they shed their angular momentum, after
which the hole is described by the Schwarzschild metric and decays
through the emission of Hawking radiation, radiating mainly into brane
modes \cite{Emparan:2000rs}.  Beyond this one needs a
full theory of quantum gravity to exactly describe the subsequent
final decay.  Because of the tension, the temperature is smaller and
the angular momentum is higher than the ``braneless'' case, and so one
would expect the lifetime of the black hole to be increased.  The
lifetime of a six-dimensional black hole can be estimated from thermodynamics as $dE/dt \sim dM/dt
\sim \left(\textrm{Area}\right)\times T_H^6$, giving $\tau \sim
1/M_6{}_\textrm{eff}\left(M/M_6{}_\textrm{eff}\right)^{5/3}$ after
integrating and using Eqs. (\ref{eqn:horizon}) and (\ref{eqn:temp}).  So
one should expect that $\tau \approx B^{-8/3}\tau_0$, where $\tau_0$
is the tensionless case.  Thus, the black holes should linger a while
longer after production.  Finally the angular velocity of
the horizon is given by
\be
\Omega_H = \frac{a_*}{r_H\left(1+a_*^2\right)},\\
\label{eqn:angvel}
\ee
and is found by looking at null geodesics in the $\phi$
direction.  The velocity is smaller than the zero tension limit since the horizon is larger.

Eq. (\ref{eqn:angvel}) has been determined for 
a black hole with angular momentum along the brane.
When 
the angular momentum has only bulk components, the hole could be rotating through the bulk angle from which the deficit angle had been cut.  The hole would be described by Eq. (\ref{eqn:aBmetric}), but with $\phi \rightarrow B\phi$, instead of rescaling $\psi$.  In this case, the angular
  velocity would read
\be
\Omega_{H} = \frac{a_*}{B\,r_H\left(1+a_*^2\right)},\\
\label{eqn:angvelphi}
\ee
which is \emph{larger} than the tensionless
  case ($\Omega_H \sim B^{-2/3} \Omega_{H \,0}$, after including the factor of
  $B^{-1/3}$ in $r_H$, and where $\Omega_{H\,
  0}$ is the tensionless case).  The increase is due to the fact that
  the hole is now only completing a rotation through $2\pi B \le 2\pi$
  radians, as discussed above, and so the velocity should be higher
  for a fixed angular momentum.  This would be an additional
  amplification due to the tension \emph{beyond} the rescaling of the Planck
  mass.  This is in contrast to Eq. (\ref{eqn:angvel}) where the
  angular velocity is \emph{reduced} due to the tension.

The black holes with their angular momentum pointing along the
brane will lose this momentum through the Penrose process
(superradiance).  This process enhances the emission of higher-spin
particles such that graviton emission could be the dominant effect
\cite{Frolov:2002gf}, though other brane fields still
participate.  After the hole has spun down and enters the Hawking
phase of its evaporation, the emission of gravitons into the bulk
could possibly cause the hole enough recoil to leave the brane
\cite{Flachi:2006ev} - \cite{Cardoso:2005mh}, overcoming the small perturbation restoring
force due to the tension.  However, for
codimension-2 branes this effect is likely to be small \cite{Gingrich:2007fk}.

Although the presence of the brane can give additional effects as
described above, it is clear that the main effect of the tension really is, \emph{in every
case so far considered}, to simply rescale $M_6^4 \rightarrow
M_6^4{}_\textrm{eff}$.  This can have useful experimental consequences.  For
example, since the cross section for production of a black hole is $\sigma \approx \pi
r_H^2$, one would expect an increase in the cross section due to the
tension on the brane \cite{Dai:2006hf}.  The black holes would be easier to make, and so
one should find an increase in the production rate.  This effect, combined with the increase in
the lifetime described above should be useful in the experimental study of
TeV-scale black holes at the LHC, if nature has chosen to behave in
this way (see \cite{Dai:2006hf} for a discussion of the effects of a nonzero tension on
the evaporation of a Schwarzschild black hole).

In fact, one can see that the tension on the brane may be able help in the
production of black holes in another way.  The LHC will be able to
create TeV-scale black holes, and it is usually assumed that the
fundamental scale of gravity needs to be in this range for this
process to occur.  However, due to the lightning rod effect, one might need
only $M_6{}_\textrm{eff}$ to be in the TeV range.  Since
$M_6{}_\textrm{eff} = B^{1/4}M_6$ one can see that for near-critical
branes one might have a value for $M_6$ higher than a TeV and still
create black holes!  Hence, the fundamental scale of gravity could potentially be
higher than the experimentally accessible range, but we still have a
possibility of exploring quantum gravity if the tension is near critical.

\section{Conclusions}

In general, braneworld systems can be very difficult to analyze.
Tension on the brane can curve the brane, as well as the bulk.
However, codimension-2 branes allow the tension to be offloaded into
the bulk, leaving the brane flat which can considerably simplify the
analysis.  In this note we have benefited significantly from this
simplification to describe the \emph{exact} solution for a black hole
spinning on a codimension-2 brane.  This solution shows the same
lightning rod behavior as the Schwarzschild solution of
\cite{Kaloper:2006ek}.  This behavior rescales the fundamental Planck
mass, amplifying various properties of the black hole
such as the event horizon, the angular momentum, and the lifetime.  The
temperature of the black hole is correspondingly smaller.  The
combination of all these effects may prove to be useful in the
experimental study of TeV scale black holes since the holes
should be easier to produce and live longer than the corresponding
tensionless case.

While the main effect of the tension on the brane is to rescale the
Planck mass, it can also have additional effects.   As we have seen in
Eq. (\ref{eqn:angvelphi}), the angular velocity of a black hole with bulk angular momentum can be amplified to
a value \emph{larger} than the corresponding tensionless case.  This effect is not present in a black hole with purely brane angular momentum components, and is an amplification beyond simply the usual lightning rod effect which
seeks to \emph{decrease} the angular velocity. 

\section{Acknowledgments}

I would like to thank my adviser, Nemanja Kaloper, for interesting and
useful discussions.

\end{document}